\documentclass[11pt]{article}

\usepackage{tikz}
\usepackage{enumerate}
\usepackage{array}

\newcolumntype{C}[1]{>{\centering\let\newline\\\arraybackslash\hspace{0pt}}m{#1}}
\usepackage{comment}
\usepackage{pgfplots}
\pgfplotsset{compat=1.17}
\usepackage{booktabs}
\usepackage{array}

\usepackage{enumitem}
\usepackage{blindtext}
\usepackage{todonotes}
\usepackage[T1]{fontenc}
\usepackage{microtype}
\usepackage{graphicx}
\usepackage{booktabs} 

\usepackage{times}
\usepackage{epsfig}
\usepackage{graphicx}
\usepackage{amsmath}
\usepackage{amssymb}
\usepackage[font=small,labelfont=bf]{caption}

\usepackage{array}
\newcolumntype{P}[1]{>{\hspace{0pt}}p{#1}}

\usepackage{algorithm}
\usepackage{algpseudocode}
\usepackage[export]{adjustbox}
\usepackage{graphicx}
\usepackage{color, soul}
\usepackage{rotating}
\usepackage{csvsimple,longtable,booktabs}
\usepackage{tabularx}
\usepackage[section]{placeins}

\usepackage{hyperref}
\usepackage{authblk}
\newenvironment{conditions*}
  {\par\vspace{\abovedisplayskip}\noindent
   \tabularx{\columnwidth}{>{$}l<{$} @{\ : } >{\raggedright\arraybackslash}X}}
  {\endtabularx\par\vspace{\belowdisplayskip}}

\begin{document}

\title{Pack and Measure: An Effective Approach for Influence Propagation in Social Networks
}

\author{Faisal N. Abu-Khzam, Ghinwa Bou Matar and Sergio Thoumi}

\affil{Department of Computer Science and Mathematics\\ 
Lebanese American University\\
Beirut, Lebanon}

\date{}
\maketitle

\thispagestyle{empty}

\begin{abstract}
The Influence Maximization problem under the Independent Cascade model (IC) is considered. The problem asks for a minimal set of vertices to serve as {\em seed set} from which a maximum influence propagation is expected.
New seed-set selection methods are introduced based on the notions of a $d$-packing and vertex centrality. In particular, we focus on selecting seed-vertices that are far apart and whose influence-values are the highest in their local communities. Our best results are achieved via an initial computation of a $d$-Packing followed by selecting either vertices of high degree or high centrality in their respective closed neighborhoods. 
This overall {\em Pack and Measure} approach proves highly effective as a seed selection method.
\end{abstract}


\section{Introduction}

How can a group of people cause a significant spread of information and influence a substantially large number of other people in a social network? 
Addressing this question, and similar ones, gave rise to the study of influence propagation and maximization in social networks \cite{easley2010networks, Kempe:2003}.
Whether the network in question is a social network or any other type of network, such as biological network, the main objective revolves around finding a number of  vertices that can maximize the spread of information or influence. Designing an algorithm for solving this problem depends on how a vertex influences its neighbors, which is commonly referred to as the \textit{diffusion model}.


Adjacent vertices in a network can influence each other depending on the link between them, which can be given with a certain ``influence probability.'' If it is a bidirectional link, then influence is mutual and each vertex is capable of ``activating'' the other. That is, if one vertex is influenced by some spread of any nature, such as a certain campaign, it is likely to influence (i.e., activate) the other with a certain given probability. On the other hand, if the link is unidirectional, then one of the two vertices can activate the other, but not vice versa. 

The criteria according to which vertex activation takes place is known as a diffusion model. The two basic diffusion models in the literature are the Linear Threshold model (LT) \cite{mark1978} where each vertex has a specific threshold that activates it and the Independent Cascade model (IC) \cite{Kempe:2003} where each active vertex $u$ can attempt once to activate each of its neighbors $v$ with a given probability $p_{uv}$. In the Independent Cascade model, the probability could be uniform in the graph or different for each edge. Other models have been proposed in the literature \cite{md2, md3, Chen2009, md5}. The approach proposed in this paper applies to any model but we mainly consider the Independent Cascade model as a case study.

Kempe et al \cite{Kempe:2003} proved that the influence maximization problem is $NP$-hard for the IC and LT models. This has motivated the design of several heuristics and approximation algorithms \cite{karine2018}. In this paper, we consider the IC model and introduce a novel heuristic that is based on how influence degrades at a certain rate as we move away from a vertex. Our heuristic computes a "$d$-packing" to ensure that a selected seed set is not selected from a single region in the network. Finally, we propose and study a measure for the effectiveness of influence maximization algorithms. 

\section{The Influence Maximization problem}

We model a social network as a graph $G=(V,E)$ that could be directed or undirected. The sets $V$ and $E$ are the vertices/nodes and edges/links of the graph, respectively. 
In this context, the vertices represent the individuals and the edges represent an existing relation between them. 
Weights could be assigned to edges to represent the probability a vertex influences its neighbor. The distance between two vertices $u,v\in V$ is the number of edges on a shortest path between them, often denoted by $d(u,v)$.
For a given vertex $v$, we denote by $N_i(v) = \{x\in V: d(v,x)=i\}$ the set of vertices at distance exactly $i$ from $v$. Since we are considering simple graphs (no loops or multi-edges), the degree of a vertex $v$ is the number of vertices at distance one from $v$, i.e., $degree(v) = |N_1(v)|$.

Let $p: E\rightarrow [0,1]$ be a given probability function defined on the set of edges of a graph $G=(V,E)$, and let $A\subset V$ be an initial set of {\em active} vertices. 
A {\em diffusion procedure} with respect to $A$ and $p$ consists of a number of {\em activation} steps. At each step we perform the following: every edge $e$ linking a vertex $u\in A$ to $v\in V\setminus A$ is either kept with probability $p(e)$ or deleted. When the edge is kept, the vertex $v$ is added to $A$. The diffusion procedure ends when all edges between active and inactive vertices have been examined.

Given a graph $G$ and a probability function on the edges as defined above, along with an integer $k$, the {\sc Influence Maximization} problem asks for a set $A$ of cardinality at most $k$ (or exactly $k$) such that the above described activation procedure can activate a maximum number of vertices. This definition was given in 
\cite{Kempe:2003} where the problem was shown to be $NP$-Hard for the IC and LT models. In this work, and following the common practice in the literature (for the sake of proper comparison), we will use a uniform (constant) probability for all edges. 

Observe that two different diffusion procedures, starting with the same initial set of active vertices, can reach different number of vertices. Although Kempe et al. describe the {\sc Influence Maximization} problem in terms of the expected number of activated vertices, this does not avoid some vagueness in the definition. We believe a simpler, and well defined, measure would be the number of steps needed to reach the entire network or some pre-defined percentage of its vertices. The speed of propagation plays an important role in the work reported in this paper.





As mentioned above, many heuristic approximation strategies have been proposed and studied. The most basic heuristic is the random heuristic that is simply based on selecting a randomly picked set of $k$ seed vertices for some given, or user-defined, integer $k$. This had been proposed in \cite{Kempe:2003} and it has been used as a baseline method. 
Perhaps the most common heuristic for influence maximization is the one based on favoring vertices of maximum degree (the {\em Maximum Degree Heuristic}), henceforth MDH.

MDH chooses as seed set the $k$ highest degree vertices in a given graph. Despite its simplicity, this approach has one of the highest influence spreads when compared to other methods.
Multiple other heuristics have been proposed and studied \cite{hs2,hs4,hs3,hs1}. Some recent models are community-based \cite{cb2,cb3,cb1}, which try to take advantage of a {\em community structure} inside social networks: most start by detecting a community in a network and proceed by using one of the known effective heuristics to select a seed set from its vertices.

On the other hand, approximation algorithms that attempt at some guaranteed ratio bound have also been proposed. Examples include the Cost Effective Lazy Forward model (CELF) of Leskovec et al. \cite{ap1} and its improved version CELF++ \cite{ap2}, the staticgreedy approximation method of Cheng et al. \cite{ap3} as well as some other methods \cite{ap5,ap4}.

\section{Proposed Heuristic Methods and Measures}

Centrality-based measures such as the MDH could result in seed sets where the vertices have a vast majority of their neighbors in common. In some cases, this could improve the influence spread, as typically very low probabilities are used (e.g. 0.01 or 0.05), so it is unlikely a vertex gets activated if it has one active neighbor only. However, if a vertex has $t$ neighbors in the seed set, then the probability of not being activated by any of its neighbors is $<(1-p)^t$. If we use $p=0.01$, which is common in the literature, then for $t\geq 69$ the vertex will be activated with probability $>0.5$. This could happen in very dense parts of the network. Here, adding one more neighbor to the seed set could be non-optimal.

To avoid the above phenomenon and secure better global spread of influence, we propose the use the notion of a ``$d$-packing'' to obtain a seed set with less (or no) intersecting neighborhoods. A set of vertices $S\subset V(G)$ is called a $d$-packing if $\forall u,v\in S, d(u,v)>d$. Informally, it is a subset of the vertices in which any two vertices are at distance greater than $d$ from each other. For $d>=2$, this guarantees that the vertices in our seed set have no overlapping neighbors. 

\subsection{d-Packing coupled with the Maximum Degree Heuristic}

Computing a $d$-packing will be used in this paper as a first step in computing a seed set. In fact, $d$-packing followed by MDH can be a more promising heuristic. This is simply performed by selecting a vertex of maximum degree from the closed neighborhood of each elements of a $d$-packing. 

We should note that we do not aim at computing a largest $d$-packing, which is $NP$-hard. Obviously, computing any maximal $d$-packing is efficient: for each selected vertex $v$, delete all vertices that are within distance $d$ from $v$. As such, one can select a vertex of maximum degree in the graph at each step, until the graph is empty. If the (sorted) degree sequence is given, this algorithm has the same running time as Breadth-First Search.


\subsection{Speed of Propagation}


Influence maximization algorithms are usually compared by fixing the seed set size and checking the number of active vertices in the graph when the spread process terminates. This does not give an insight on the speed of propagation, which could be an essential factor when choosing among two different seed sets with similar influence spread. To measure the speed of propagation, we can  use $p=1$: a vertex will always influence all of its neighbors. Measuring influence can either be based on the number of steps it takes for the seed set to activate all vertices in the network or on the pace at which vertices get activated. The number of steps is the minimum of shortest paths between each vertex of seed set and the last vertex that gets activated, i.e. if vertex $x$ is the last vertex to be activated then the number of steps is:
\begin{equation}
    \min_{s\in seed\, set} d(x,s)
\end{equation}

The Independent Cascade problem, along with the speed of propagation measure is not new. It as been given several names in the literature, but perhaps it was first been proposed in \cite{hopcroft2001introduction} as the Firehouse problem. However, the partial version of the problem is possibly the closest to the diffusion models that are based on low edge probabilities. This is formally defined as follows.

\vspace{10pt}
\noindent
{\sc Partial Firehouse Problem}\\
\underline{Given:} a graph $G=(V,E)$, a numer $t\in (0,1]$ along with integers $d$ and $k$;

\noindent
\underline{Question:} Is there a set $S$ of at most $k$ vertices such that at least $t|V|$ vertices of $G$ are within distance $d$ from $S$?

\vspace{10pt}

The problem is $NP$-hard and $W[2]$-hard (the Dominating Set problem is a special case; when $d=t=1$). Typical heuristic methods are based on favoring vertices of maximum degree, or maximum utility: largest number of uncovered vertices; assuming a vertex is covered when it is already within distance $d$ from a (potential) solution.

In this paper, we will not use the speed of propagation as a stand-alone measure. Instead, we couple it with the common diffusion models, which use small edge probabilities. As such we shall measure both the number of activated vertices and the speed of propagation, simply by counting the steps until no more edges can be explored via the IC diffusion model.

\subsection{The Diminishing Influence Heuristic}

When speed of propagation is the most important measure, and not the size of a solution (seed-set) $S$, a heuristic algorithm should favor the vertex that can cover a larger number of vertices that are within distance $d$ from it. With this in mind, we adopt the following general formula for measuring the influence of a vertex in a graph. We shall refer to it as the {\em diminishing influence function}:

\begin{center}
$Influence(v) = \sum a_i \hspace{1mm} w(v)$
\end{center}

\noindent where $a_i$ is a decreasing sequence that is used to model the influence of a vertex as we go farther from it. Here, we use the geometric sequence $a_k = 
\frac{1}{2^k}$. The weight function ($w$ in the above formula) assigned to a vertex in our model is assumed to be its {\em local influence}. In this work, and since the edge probability function is assumed to be constant, we calculate the influence of a vertex as:

\begin{equation}
   Influence(v) = \sum_{i=1}^{l} \frac{1}{2^i} |N_{i}(v)|, \text{where }  l \leq \max_{{w \in V}} d(v,w)
\end{equation}

Using the above influence measure, we propose the following (corresponding)  {\em Diminishing Influence Heuristic} algorithm.

\vspace{5pt}

\begin{algorithm}
\caption{Diminishing Influence Heuristic}
\begin{algorithmic}[1]
\For{each vertex $v$ in $G$}

    \State $distancefrom=BFS(G,v)$
    \State {\sc Calculate}$(G, v, distancefrom)$
\EndFor

\Function{Calculate}{$G, vertex, distancefrom$}
    \For{each $othervertex$ in $G$}
        \State $value[vertex] = value[vertex] + \frac{1}{2^{distancefrom[othervertex]}}$
    \EndFor
\EndFunction
\end{algorithmic}
\end{algorithm}

The $distancefrom$ array contains the distance between the current vertex and all other vertices in the graph. 
The running time of the above algorithm is in $O(n^2)$.

Finally, and to secure fast coverage as noted above, one can start by computing a $d$-packing, and then favor the selection of vertices of highest diminishing influence function that is close to each element of the $d$-packing. This guarantees a set of ``scattered'' seed elements of potentially high influence. 

\vspace{5pt}

\begin{algorithm}
\caption{Diminishing Influence Heuristic + $d$-packing}
\begin{algorithmic}[1]

\State $candidates = d$-$packing (G, k, d)$

\For{each vertex $v$ in $candidates$}

    \State $distancefrom=BFS(G,v)$
    \State {\sc Calculate}$(G, v, distancefrom)$
\EndFor

\Function{Calculate}{$G, vertex, distancefrom$}
    \For{each $othervertex$ in $G$}
        \State $value[vertex] = value[vertex] + \frac{1}{2^{distancefrom[othervertex]}}$
    \EndFor
\EndFunction
\end{algorithmic}
\end{algorithm}

The $d$-packing function in the above algorithm returns a set of $k$ vertices any two of which are at distance greater than $d$ from each other. The effect of applying the above algorithms is studied, empirically, in the next section.

\section{Experimental Results}

We conducted our experiments on the ca-GrQc and ca-AstroPh collaboration networks from the SNAP dataset \cite{snapnets}. Their properties can be found in Table \ref{tab:properties} . We note that $d$ should be tuned based on the given network. Obviously, the best value used on one network might not guarantee the best result on a different one. The results reported below present the average values computed based on 1000 iterations, rounded to the nearest integer.

\vspace{5pt}
    
\begin{table}[H]
\centering
\begin{tabular}{|c|c|c|}
\hline
\textbf{Network} & \textbf{Vertices} & \textbf{Edges} \\ \hline
ca-GrQc & 5242 & 14496 \\ \hline
ca-AstroPh & 18772 & 198110 \\ \hline
\end{tabular}%
\caption{Number of vertices and edges for each graph.}
\label{tab:properties}
\end{table}

\begin{table}[H]
\centering
\resizebox{\columnwidth}{!}{%
\begin{tabular}
{|c|c|c|c|c|c|}
\hline
\textbf{Seed Set Size} & \textbf{Maximum Degree} & \textbf{Maximum Degree + 2-packing} & \textbf{Diminishing Influence + 2-packing} \\ \hline
10 & 30 & 26 & 23 \\ \hline
20 & 45 & 42 & 39 \\ \hline
30 & 54 & 56 & 51 \\ \hline
40 & 59 & 69 & 66 \\ \hline
50 & 68 & 82 & 78 \\ \hline
\end{tabular}%
}
\caption{Vertices reached in ca-GrQc with $p=0.01$.}
\label{tab:reachability_values_p001}
\end{table}

\noindent The improvement is more apparent when using slightly larger probabilities such as $p=0.05$.

\begin{table}[H]
\centering
\resizebox{\columnwidth}{!}{%
\begin{tabular}{|c|c|c|c|c|c|}
\hline
\textbf{Seed Set Size} & \textbf{Maximum Degree} & \textbf{Maximum Degree + 2-packing} & \textbf{Diminishing Influence + 2-packing} \\ \hline
10 & 200 & 364 & 338 \\ \hline
20 & 250 & 438 & 406 \\ \hline
30 & 253 & 486 & 443 \\ \hline
40 & 251 & 515 & 493 \\ \hline
50 & 275 & 549 & 519 \\ \hline
\end{tabular}%
}
\caption{Vertices reached in ca-GrQc with $p=0.05$.}
\label{tab:reachability_values_p005}
\end{table}

\noindent
In some cases, larger values of $d$ do not increase the number of vertices reached, however, they speed-up the propagation process. For ca-AstroPh, we present the $d$-packing results for $d=1$ and $d=4$. We note that 1-packing reaches the most vertices, but it does not improve the speed of propagation. Interestingly, 4-packing does not increase the number of nodes reached, but it offers faster propagation (as shown in the next section). 

\vspace{5pt}

\begin{table}[H]
\centering
\resizebox{\columnwidth}{!}{%
\begin{tabular}{|C{2.5cm}|C{2.5cm}|C{2.5cm}|C{2.5cm}|C{2.5cm}|C{2.5cm}|}
\hline
\textbf{Seed Set Size} & \textbf{Maximum Degree} & \textbf{Maximum Degree + 1-packing} & \textbf{Maximum Degree + 4-packing} & \textbf{Diminishing Influence + 1-packing}  & \textbf{Diminishing Influence + 4-packing}\\ \hline 
10 & 1754 & 1774 & 1741 & 1776 & 1731 \\ \hline
20 & 1760 & 1793 & 1757 & 1803 & 1751 \\ \hline
30 & 1766 & 1821 & 1775 & 1838 & 1756 \\ \hline
40 & 1783 & 1856 & 1787 & 1867 & 1778\\ \hline
50 & 1792 & 1874 & 1790 & 1899 & 1785 \\ \hline
\end{tabular}%
}
\caption{Vertices reached in ca-AstroPh with $p=0.01$.}
\label{tab:astroph}
\end{table}

\subsection{Networks with Scattered Dense Communities}

Real-life social networks are usually formed of scattered dense communities. Surprisingly, this is not reflected in the current networks used in the literature (e.g. citation networks). The SNAP dataset collection provides networks containing different communities. However, these are not usually used, most likely because they are too large. Centrality-based heuristics such as the MDH would not perform well in networks with scattered dense communities as the chosen seed set would be almost entirely from the densest community. Thus, other communities would not be activated/influenced. Of course, the assumption is that communities are loosely connected and low probabilities are typically used. In such networks, using $d$-packing alongside the chosen heuristic provides substantial improvements. To test the propagation of influence in such cases, and as preliminary experiments, we generated a network by creating four highly dense regions (cliques) of varying sizes, then we connected each clique to two others via simple, relatively short, paths. The resulting network has 1586 vertices and 318015 edges, with the largest clique having 500 vertices. Our results are presented in Table \ref{tab:4cliques_reachability_values_p001}.

\vspace{5pt}

\begin{table}[H]
\centering
\resizebox{\columnwidth}{!}{%
\begin{tabular}{|C{3cm}|C{3cm}|C{3cm}|C{3cm}|C{3cm}|C{3cm}|}
\hline
\textbf{Seed Set Size} & \textbf{Maximum Degree} & \textbf{Maximum Degree + 9-packing} & \textbf{Diminishing Influence} &\textbf{Diminishing Influence + 9-packing} \\ \hline
10 & 496 & 1472 & 496 & 1456 \\ \hline
20 & 496 &  1462 & 496 & 1465 \\ \hline
30 & 496 & 1466 & 496 & 1459\\ \hline
40 & 496 & 1463 &  496&  1461\\ \hline
50 & 496 & 1461  & 496 &  1462 \\ \hline
\end{tabular}%
}
\caption{Vertices reached in the synthetic network with $p=0.01$.}
\label{tab:4cliques_reachability_values_p001}
\end{table}

Without a $d$-packing, all the vertices in the seed set were chosen from the densest community and their influence did not propagate well to other communities in the network. Since each community is very dense, choosing another vertex from it becomes non-optimal at a certain point, as explained in Section 3.
We should note that more work is currently underway on networks with scattered dense communities, so the above table shows (notably promising) preliminary results  at the time of writing this paper.

\subsection{Speed of Propagation}

For comparison based on the speed of propagation, we used the same values of $d$ as above. Higher values could potentially provide faster propagation but the seed set might reach less vertices, depending on the network structure. However, in this case the main objective is to provide a single seed set that reaches a high number of vertices with a fast speed of propagation. The below table shows the results for our proposed measure. As expected, seed sets selected via $d$-packing cover the network faster. Most importantly, the Diminishing Influence Heuristic proved to be better in this case, especially 
on the synthetic network that models  scattered dense communities. While these results are promising, more experiments should be conducted on networks with scattered dense communities. 

\vspace{5pt}

\begin{table}[H]
\resizebox{\columnwidth}{!}{%
\begin{tabular}{|c|c|c|c|c|c|}
\hline
\textbf{Seed Set Size} & \textbf{Maximum Degree} & \textbf{Maximum Degree + 2-packing} & \textbf{Diminishing Influence + 2-packing} \\ \hline
10 & 10 & 9  & 8 \\ \hline
20 & 10 & 9  & 8 \\ \hline
30 & 9 & 8 & 8 \\ \hline
40 & 9 & 8  & 8 \\ \hline
50 & 9 & 8 & 8 \\ \hline
\end{tabular}%
}
\caption{Number of steps needed to cover the ca-GrQc network.}
\label{tab:steps}
\end{table}

\begin{table}[H]
\resizebox{\columnwidth}{!}{
\begin{tabular}{|C{2.5cm}|C{2.5cm}|C{2.5cm}|C{2.5cm}|C{2.5cm}|C{2.5cm}|}
\hline
\textbf{Seed Set Size} & \textbf{Maximum Degree} & \textbf{Maximum Degree + 1-packing} & \textbf{Maximum Degree + 4-packing} & \textbf{Diminishing Influence + 1-packing}  & \textbf{Diminishing Influence + 4-packing}\\ \hline
10 & 8 & 8 & 8 & 9 & 8 \\ \hline
20 & 8 & 8 & 7 & 9 & 7\\ \hline
30 & 8 & 8 & 7 & 8 & 6 \\ \hline
40 & 8 & 8 & 6 & 8 & 6\\ \hline
50 & 8 & 8 & 6 & 8 & 6\\ \hline
\end{tabular}%
}
\caption{Number of steps needed to cover the ca-AstroPh network.}
\label{tab:steps2}
\end{table}

\begin{table}[H]
\centering
\resizebox{\columnwidth}{!}{%
\begin{tabular}{|c|c|c|c|c|c|}
\hline
\textbf{Seed Set Size} & \textbf{Maximum Degree} & \textbf{Maximum Degree + 9-packing} &\textbf{Diminishing Influence + 9-packing} \\ \hline
10 & 22 &  6 &  5\\ \hline
20 &  22& 6 & 5 \\ \hline
30 & 22 & 6 & 5 \\ \hline
40 & 22 & 6  & 5 \\ \hline
50 & 22 & 6  & 5 \\ \hline
\end{tabular}%
}
\caption{Number of steps needed to cover the synthetic network.}
\label{tab:steps4c}
\end{table}

Finally, it should be noted that $d$-packing with diminishing influence tends to behave much better on networks with larger edge probabilities, but we restricted our presentation to low edge probabilities, being the most used in the literature.

\section{Conclusion and Future Work}

Several algorithms have been proposed for computing what is known as a seed set, through which influence propagation in a network is possibly maximized. In this paper, we introduced a method that is based on computing a $d$-packing prior to applying any heuristic method. This is simply motivated by the fact that a $d$-packing is nothing but a set of ``scattered'' vertices in a network.

We further proposed a heuristic method that takes into consideration (i) the pace of influence reduction of a vertex as we gradually move away from it, and (ii) the fact that some individuals can  have greater influence if they have a possibly small number of links, but many of these links are to people who happen to have a relatively large number of links. Our ``diminishing influence'' measure is based on the use of a suitable decreasing sequence. We used a particular geometric sequence that halves the influence per step, but one can explore the use of other decreasing sequences that might potentially depend on the the network structure.

A thorough experimental study showed the notable effectiveness of the proposed $d$-packing approach. In all experiments, computing a $d$-packing followed by the popular maximum-degree heuristic exhibited a better network coverage. Furthermore, $d$-packing coupled with our diminishing influence heuristic proved to be more effective on networks with what we called ``scattered dense communities.'' Investigating this type of networks is an ongoing research.

The use of a $d$-packing prior to applying heuristic methods can potentially give better results when considering/solving 
other 
related social network problems, such as {\sc Harmless Set} \cite{abu_harmlesset,
BazganC14, Chen2009} and {\sc Positive Influence Dominating Set} problem \cite{karine2018, Chen2009, RaiePids}. Another potential future direction would be to explore the same notions on dynamic networks, where links can be added or deleted after a ``satisfactory'' seed set is computed. Parameterized dynamic problems might also be explored in the same context \cite{abu2015}.

Finally, we proposed a possibly enhanced measure of propagation by using the speed of propagation along with both the LT and IC models, simply by counting the number of steps needed to reach/explore all the vertices in the given network. As far as we know, this natural model has not been used before in the study of influence propagation in social networks. The use of this measure shows a potential advantage of our diminishing influence heuristic, again when preceded by the computation of a suitable $d$-packing.
A possibly better way of using the speed of propagation measure would be in computing the rate at which activation varies (increases or decreases) with time as diffusion moves away from the seed set.


\bibliographystyle{abbrv}
\bibliography{References}

\end{document}